\documentclass[11pt,a4paper]{article}
\usepackage[utf8]{inputenc}

\usepackage[normalem]{ulem}
\usepackage{hyperref}
\usepackage{authblk}
\usepackage{color}
\usepackage{fullpage}

\usepackage{graphicx}
\usepackage{amsfonts}
\usepackage{amsmath,bm}
\usepackage{enumitem}
\usepackage[linesnumbered,ruled,vlined]{algorithm2e}

\usepackage{moreverb,url}

\title{Systematic errors in estimates of $R_t$ from symptomatic cases in the presence of observation bias}
\author{Guido Sanguinetti$^{1,2}$}

\affil{\footnotesize$^{1}$SISSA, Trieste, Italy\\
$^{2}$School of Informatics, University of Edinburgh, UK\\ gsanguin@sissa.it}
 \date{23 November 2020}
\begin{document}
\maketitle
\begin{abstract}
    We consider the problem of estimating the reproduction number $R_t$ of an epidemic for populations where the probability of detection of cases depends on a known covariate. We argue that in such cases the normal empirical estimator can fail when the prevalence of cases among groups changes with time. We propose a Bayesian strategy to resolve the problem, as well as a simple solution in the case of large number of cases. We illustrate the issue and its solution on a simple yet realistic simulation study, and discuss the general relevance of the issue to the current covid19 pandemic.
\end{abstract}

\section{Introduction}
The instantaneous reproduction number $R_t$ is an effective empirical measurement of the velocity with which an epidemic is spreading through a population. A value of $R_t$ greater than 1 denotes an epidemic which is growing exponentially at that time, while values smaller than 1 are witnessed in declining phases of the epidemic. Because of its immediate interpretation and its independence of detailed modelling assumptions, estimate of the value of $R_t$ are of immense importance to public health experts and policy makers, and are normally an essential consideration in determining measures to fight the spread of an epidemic.

$R_t$ is defined as the expected number of secondary infections at time $t$ from each infected case, or equivalently the ratio of the number of new infected cases at time $t$ to the average infectiousness of individuals at that time (see Section \ref{maths} for a more precise definition). As such, it returns a local (in time), model-free exponential approximation to the epidemic dynamics, providing precious information to population health scientists and policy makers. While alternative approximations which do not assume exponential dynamics exist \cite{chowell2016characterizing}, in practice most epidemiological studies utilise the exponential approximation approach, as witnessed in the manifold studies of the dynamics of the recent covid19 pandemic, e.g. \cite{flaxman2020report,kwok2020herd}.  Because the ascertainment of dates of infection is in general difficult if not impossible, in practice $R_t$ is estimated on the basis of new cases whose symptom onset happens at time $t$, as suggested in several publications, see e.g. \cite{cori2013new}. Therefore, estimates of $R_t$ are describing the dynamics of the epidemic in the subset of the infected population which develops sufficiently severe symptoms as to be detected as infected. Such a restriction is often considered as a positive feature, as it avoids the unavoidable uncertainty linked to detection of asymptomatic and paucisymptomatic cases, which is heavily confounded by testing strategies which may be highly time-varying.

Mathematically, computing $R_t$ from symptomatic cases only is fully justified under the assumption that new symptomatic cases constitute a constant fraction $\alpha\le 1$ of the total (unknown) number of new cases in a day. Such an assumption guarantees that the overall dynamics of the epidemic among the symptomatic cases are identical to the true epidemic, just rescaled by a factor $\alpha$. However, empirical evidence from the covid19 pandemic in western Europe suggests that this assumption is not always justified.

It is well known that infection by the new coronavirus results in a large fraction of asymptomatic cases, however such fraction is heavily dependent on a number of additional covariates. In particular, it is well known that younger people are far less likely to develop symptoms than older people \cite{davies2020age}, therefore if the progression of the disease is different between different age groups, it is plausible that the ratio $\alpha$ of symptomatic cases to total cases will not remain constant, but it will follow the spread of the epidemic across the age groups. Such a situation has been widely observed during the autumnal second wave of the infection, when the virus spread rapidly among younger people in holiday resorts in the summer, and gradually started spreading to older people during the autumn, with a gradual increase of the average age of the cases detected witnessed.

In this brief paper, I show that estimating $R_t$ from symptomatic cases only can lead to mistaken conclusions when the probability of developing symptoms depends on a covariate which is not constant in time. I also show how the problem can be fixed by introducing latent variables, and how an extremely simple correction can provide very accurate maximum likelihood estimates of the true $R_t$ when the number of cases is large and the dependence of the probability of developing symptoms on the covariate is known. Numerical examples on a simple yet informative model confirm the potential relevance of these considerations.

\section{Mathematical formulation of the problem and proposed solutions}\label{maths}
\subsection{Definitions and problem statement}
The instantaneous reproduction number $R_t$ is defined mathematically by the equation \begin{equation}
    I_t=R_t\sum_{\tau=1}^tp(\tau)I_{t-\tau}\label{Rtdef}
\end{equation}
where $I_t$ is the number of infected individuals at time $t$, and $p(\tau)$ is the distribution over generation times, effectively accounting for incubation times. In practice, the distribution over generation times is assumed known from epidemiological studies; for example, the Italian Istituto Superiore di Sanit\'a (ISS) uses estimates of generation times distribution obtained from the early Covid19 outbreak in Lombardy \cite{cereda2020early} to compute $R_t$ estimates in its weekly monitoring. Equation \eqref{Rtdef} defines an instantaneous rate of change of the infected population independently of the stochastic generative process underpinning the epidemic. Practically one simply combines equation \eqref{Rtdef} with a Poisson or Gaussian noise distribution on the output $I_t$ to provide straightforward Bayesian or maximum likelihood estimates of the parameter of interest $R_t$.

Equation \eqref{Rtdef} is linear in the number of new daily infections $I_t$, therefore it still holds when the vector of new infections is rescaled by a constant. Therefore, under the assumption that symptomatic cases are a constant fraction of new cases, estimates of $R_t$ obtained using only symptomatic cases will be identical to estimates obtained using all new infections. Such an assumption will however fail if the fraction of symptomatic cases depends on time. 

The Covid19 pandemic in Western Europe, during its second wave in the autumn of 2020, has shown a constant gradual increase in the median age of cases. As an example, the Italian ISS reported a median age of Covid19 positives of 29 in the period 17-23 August 2020, which increased to 47 in the period from mid October to mid November 2020\footnote{Weekly data on median age of cases appear to be unavailable after September. Data from ISS reports \url{https://www.epicentro.iss.it/coronavirus/sars-cov-2-dashboard}.}. The fraction of Covid19 cases presenting clinically relevant symptoms is estimated to vary from less than 20\% in young individuals to over 70\% in elderly people \cite{davies2020age}. Therefore, the symptomatics rate must have changed in the period August-November, invalidating the assumptions underlying the procedure to estimate $R_t$. In particular, part of the increase in symptomatic cases will be the result of an increased rate of symptomatic cases in older individuals, leading therefore to an inflation in the estimates of $R_t$.
\subsection{Latent variable model}
A natural solution to the problem is to retain the original definition of $R_t$ in terms of infected numbers, and to introduce observable variables $S_t(l)$ denoting number of symptomatic cases at time $t$ with covariate value equal to $l$\footnote{This treatment is equally suited to handling discrete or continuous covariates, however for practical convenience it might be easier to group continuous covariates such as age into discrete classes.}. The observables are obtained from the latent number of infected cases with covariate $l$ $I_t(l)$ through a Binomial observation model\begin{equation}
    S_t(l)\sim\mathrm{Binom}\left(I_t(l),\pi_l\right)
\end{equation}
where $\pi_l$ is the symptomatic rate for infected cases in group $l$. Assuming this rate to be known (for example from epidemiological studies such as \cite{davies2020age}), then  the variables $I_t(l)$ are all independent {\it a posteriori}\footnote{This is because the renewal equations  \eqref{Rtdef} and \eqref{RtAges} are deterministic conditioned on $I_t$ and $I_t(l)$ respectively.} and, given a suitable prior, they can be estimated independently via e.g. the Metropolis-Hastings (M-H) algorithm or any other Bayesian sampler\footnote{I am not aware of a conjugate prior distribution over the number of trials to a binomial likelihood where the success rate is known.}. To obtain a posterior distribution over the parameter of interest $R_t$, one simply needs to rewrite equation \eqref{Rtdef} by summing over the covariate values \begin{equation}
    \sum_l I_t(l)=R_t\sum_{\tau,l}\pi_lp(\tau)I_{t-tau}(l)\label{RtAges}
\end{equation}
and plugging samples of $I_t(l)$ in equation \eqref{RtAges}\footnote{We assume that generation times are independt of the covariate $l$.}.

If the symptomatic rate $\pi_l$ is not known, then it can also be estimated within a Bayesian framework. An efficient solution could be to assign each $\pi_l$ independent Beta priors and run a block Metropolis-within-Gibbs scheme: conditioned on observables and $\pi_l$ values, the $I_t(l)$ can be sampled in parallel via independent M-H samplers. Given sampled values of $I_t(l)$, the posterior on each $\pi_l$ value can be obtained analytically and sample values of $\pi_l$ could be drawn in a straightforward way. It is however probable that informative priors on $\pi_l$ would be needed for the estimates to have an acceptable level of uncertainty.
\subsection{Gaussian approximation and analytical corrections}
Assuming that the numbers of infected individuals for each group $I_t(l)$ are sufficiently large, and that the symptomatic rates $\pi_l$ are also not too close to zero or one, it is then possible to approximate the binomial distribution with a Gaussian conditional on $I_t(l)$, so that \begin{equation}
    p\left(S_t(l)\vert I_t(l),\pi_l\right)\simeq\mathcal{N}\left(\pi_l I_t(l),\pi_l(1-\pi_l) I_t(l)\right).\label{GaussApp}
\end{equation}
Introducing $\delta_t(l)=I_t(l)-\frac{S_t(l)}{\pi_l}$, we can rewrite the exponent in \eqref{GaussApp} as \[
-\frac{\left(S_t(l)-\pi_l I_t(l)\right)^2}{\pi_l(1-\pi_l) I_t(l)}=-\frac{\pi_l}{1-\pi_l}\frac{\delta_t(l)^2}{\frac{S_t(l)}{\pi_l}\left(1+\frac{\pi_l\delta}{s_t(l)}\right)}
\]
from which it is clear that the maximum likelihood value for $I_t(l)$ is obtained for $\delta=0$, i.e. when $I_t(l)=\frac{S_t(l)}{\pi_l}$. Plugging this estimator in the equation for $R_t$ \eqref{RtAges}
yields the following estimator\begin{equation}
    R_t^{MLE}=\frac{\sum_lS_t(l)\pi_l^{-1}}{\sum_{\tau=1}^t\sum_lS_{t-\tau}(l)\pi_l^{-1}p(\tau)}\label{Rt_MLE}
\end{equation} 
\section{Numerical illustration}
To illustrate the impact of violating the assumption of constant probability of symptoms onset, we analyse a simple epidemic model which simulates the spread of a disease among two equal sized populations with weak interactions. The two populations have equal size of 200.000 individuals, and differ only in the probability that an infected individual will develop symptoms: infected individuals in one population, Y, will develop symptoms with probability 0.3, while individuals in the other, O, will develop symptoms with probability 0.8. These values were chosen to mimic the values reported for COVID19 in young and old individuals \cite{davies2020age}.

\begin{figure}
    \centering
    \includegraphics[width=0.45 \textwidth]{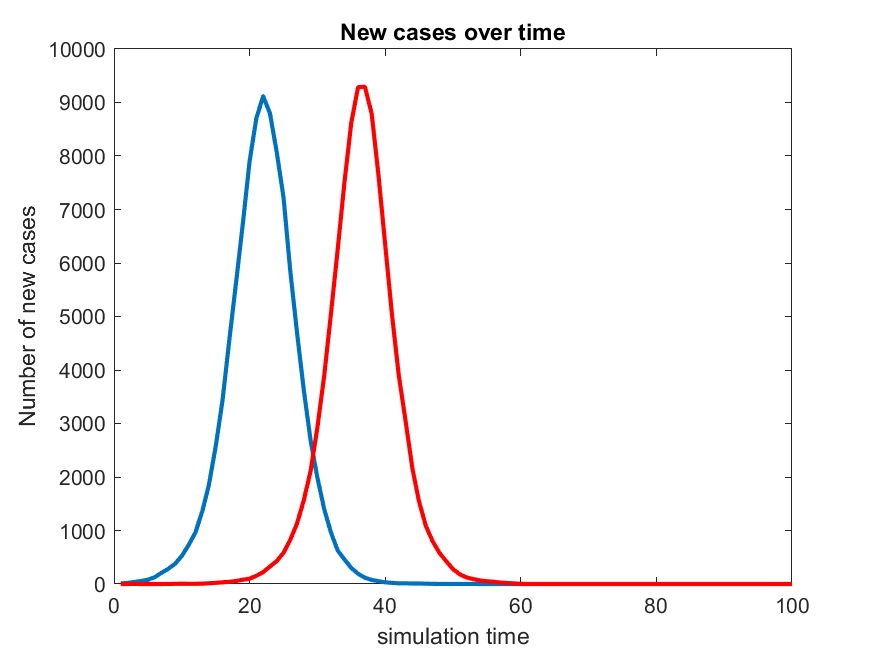}
    \includegraphics[width=0.45 \textwidth]{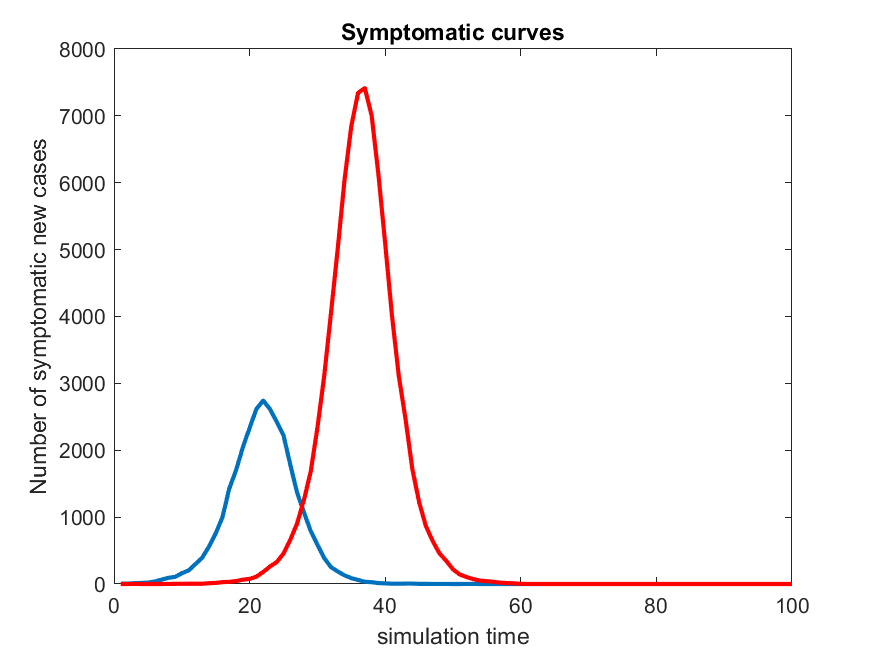}\\
    
    \includegraphics[width=0.45 \textwidth]{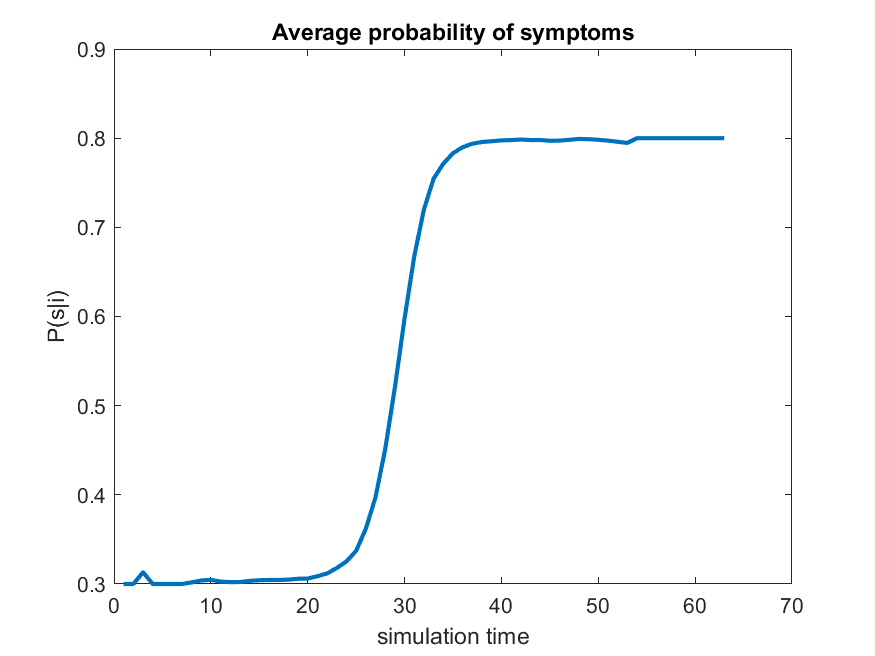}
    \includegraphics[width=0.45 \textwidth]{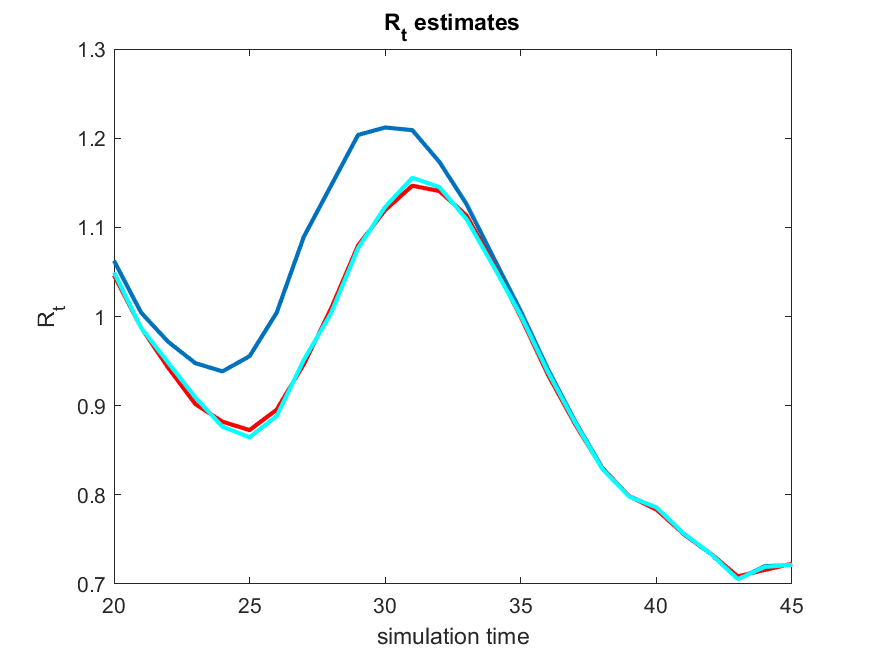}
    \caption{Example results in a simple SI epidemic model with two populations with different probabilities of symptoms onset (young, $p(s_y\vert I_y)=0.3$, and old $p(s_o\vert I_o)=0.8$): top left, new daily infections for young (blue) and old (red) populations. An early wave of infection in the young population slowly trickles to the old population, which develops a nearly identical wave approximately 20 days later. Top right: observed new daily symptomatic cases, showing an apparently much larger infection among the older. Bottom left, fraction of new symptomatic cases over total new cases, showing a clear rapid increment when the infection among the old population starts to dominate. Bottom right, estimated $R_t$ using symptomatic cases (blue), total cases (red) and corrected maximum likelihood estimates (cyan), in the time interval when the fraction of symptomatic cases changes. Standard estimates based on symptomatic cases only overestimate $R_t$ by approximately 10\%, while the corrected maximum likelihood estimate is nearly identical to the true $R_t$ value consistently throughout the period.}
    \label{fig:sim_model}
\end{figure}

Infection dynamics are almost identical in the two sub-populations, according to a stochastic SI model with $R_0=1.4$. To simplify further, we assumed that infected individuals remain infectious for only 1 time step, i.e. there is no incubation period and can only infect individuals the day after they were infected. Individuals in the Y population can also infect with low probability individuals in the O population. The resulting epidemic dynamics over 100 simulation steps are shown in Figure \ref{fig:sim_model} top left, resulting in two nearly identical waves of infections with a delay of approximately 20 time units. The observed dynamics of symptomatic cases is however very different, with a first low peak of symptomatic Y cases and a later, much larger peak of O cases (Fig. \ref{fig:sim_model} top right). The probability of showing symptoms transitions sharply at around time 30 from the Y probability (0.3) to the O probability (0.8), reflecting the change in prevalence of the infection among the two populations (Fig \ref{fig:sim_model} bottom left). Correspondingly, around time 30 we see a significant deviation of the $R_t$ estimate computed on symptomatics only (Fig \ref{fig:sim_model} bottom right blue line) from the ground truth estimate computed on all cases (red line). The Gaussian MLE correction proposed in equation \eqref{Rt_MLE} is in almost perfect agreement with the true $R_t$ (cyan line).

 While the analysis of this simple model reveals a marked deviation of the estimated $R_t$ from the true $R_t$, the effect appears relatively modest, approximately around 10\% relative error. One possible explanation for this is that the simple model does not involve incubation times, so that the rate of change of the fraction of symptomatics is slow relative to the fast contagion time.
 
 I therefore explored a different scenario where the generation time distribution is zero everywhere except for 0.5 at days -3 and -4 (i.e., there is an incubation time of 3 days and infectiousness lasts two days). I set the $R_0$ to 2.4 to yield similar epidemic dynamics to the simple case in Figure \ref{fig:sim_model}. The deviation of the estimates from symptomatic cases from the true values appear more marked, presumably because the nonzero incubation time leads to a sharper apparent increment in symptomatic cases due to the shift in population. Once again, the simple correction from the Gaussian approximation seems to rescue the problem to a large extent.
 
 Quantifying the errors in estimates across 100 independent simulations, we see that for both the simple model with no incubation and the more complex model the Gaussian approximation yields a very significant improvement in estimates of the true $R_t$, with error reduction ranging from a factor 7 in the simple model to over an order of magnitude for the more complex model.
 
 \begin{figure}
     \centering
     \includegraphics[width=0.45\textwidth]{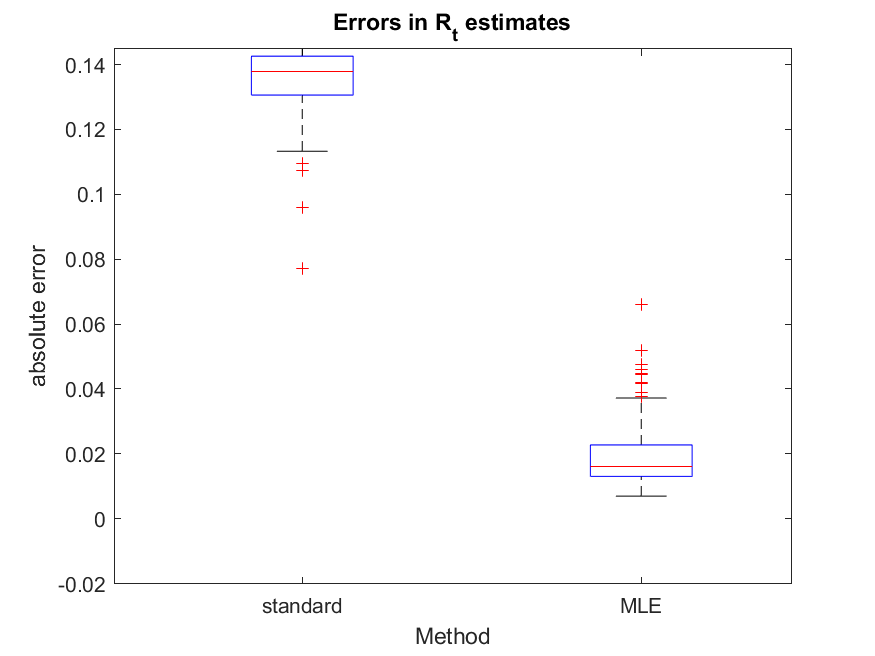}
     \includegraphics[width=0.45\textwidth]{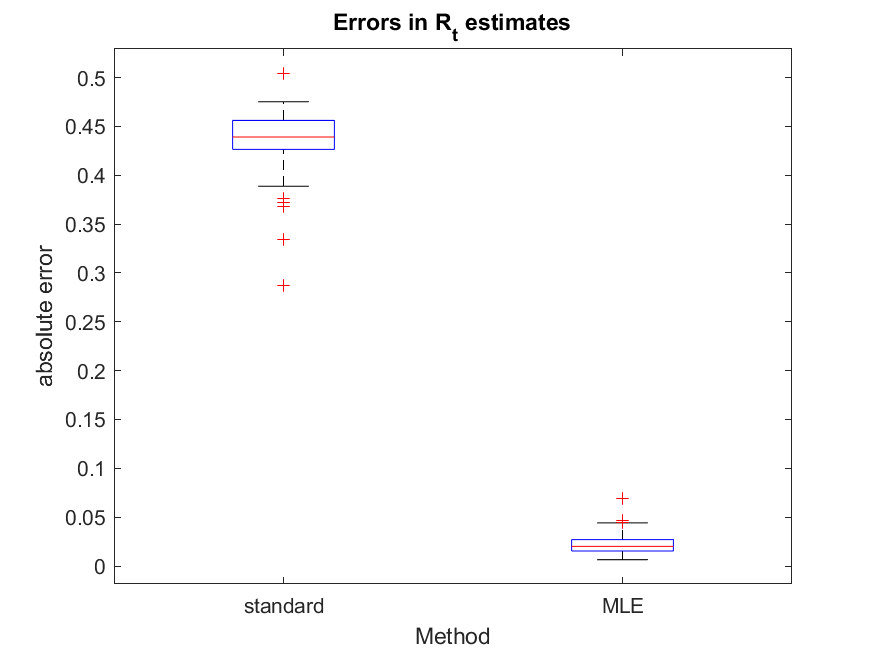}
     \caption{Errors in estimates of $R_t$ over 100 independent simulations: {\it left} boxplot of differences in estimates from symptomatics only (left box) and corrected by Gaussian approximation (right box) in model without incubation time; {\it right} similar as left panel but on model with incubation time}
     \label{fig:my_label}
 \end{figure}

\section{Discussion}
In this short paper, I have identified and discussed a possible shortcoming in the current method for estimating the reproduction number $R_t$ of an epidemic. While estimating $R_t$ from symptomatic cases only is certainly a practical solution to avoid the difficulties in estimating total infection numbers in epidemics with many asymptomatic individuals, I argue both theoretically and numerically that this strategy can yield erroneous results when the proportion of symptomatic cases is not constant in time. I also showed that the problem can be rescued to a large extent when the change of symptomatic cases is the result of a shift of the epidemic across populations with different (known) characteristics.

A major limitation of this work is the assumption that the time dependence of the fraction of symptomatics results only from the population structure and how the epidemic spreads across the population. In reality, myriad effects might lead to even larger errors, from local difficulties in recording and reporting symptomatic cases due to health services being overwhelmed, to changes in testing policies (for example leading to the identification of more or less individuals with light symptoms). Nevertheless, I would argue that, despite its limitations, the proposed analysis has practical merit, since changes in median age of infected individuals have been recorded widely in the current COVID19 pandemic, leading to errors that could easily be corrected. It is worth pointing out that even fractional errors in estimates of $R_t$ can have huge practical consequences, since the reproduction number is the main quantitative indicator adopted by governments in deciding containment measures which have frequently enormous economic impact. This issue is even more pressing because frequently estimates of $R_t$ at local level are adopted, where factors such as infections spreading in care homes (which have by definition a high median age and hence a large fraction of symptomatics) can completely dominate the recorded numbers of symptomatic cases.

Methodologically, this work points to a more substantial usage of latent variable models in epidemiological studies. Latent variable models are a popular and well established area of research in statistics and machine learning; to my knowledge, the proposal of using latent variables to estimate $R_t$ in the current covid19 pandemic has been explored in \cite{de2020impact}, where however the focus was reconstructing total numbers of infected individuals from total numbers of positive tests. It is widely believed that machine learning methods might greatly help monitoring and predicting the spread of pandemics, however their application is critically limited by data availability. Regrettably, relevant real data to test models is scarce: in the case of this work, while median age of cases and total recorded cases are widely available, I have not been able to obtain real data on symptomatic cases per day and their age categories, therefore preventing a more meaningful evaluation of the proposed methodology.

\footnotesize
\bibliographystyle{plain} 
\bibliography{covidRt}
\end{document}